# OXIDATION AND MAGNETIC STATES OF CHALCOPYRITE $CuFeS_2$: A FIRST PRINCIPLES CALCULATION

V.V. Klekovkina*, R.R. Gainov*, F.G. Vagizov*, A.V. Dooglav*, V.A. Golovanevskiy**, I.N. Pen'kov*

*Kazan Federal University, 420008 Kazan, Russia

**Curtin University, U1987 Perth, Australia

**Abstract**

The ground state band structure, magnetic moments, charges and population numbers of electronic shells of Cu and Fe atoms have been calculated for chalcopyrite $CuFeS_2$ using density functional theory. The comparison between our calculation results and experimental data (X-ray photoemission, X-ray absorption and neutron diffraction spectroscopy) has been made. Our calculations predict a formal oxidation state for chalcopyrite as $Cu^{1+}Fe^{3+}S_2^{2-}$. However, the assignment of formal valence state to transition metal atoms appears to be oversimplified. It is anticipated that the valence state can be confirmed experimentally by nuclear magnetic and nuclear quadrupole resonance and Mössbauer spectroscopy methods.

**Introduction**

Chalcogenide materials of $Cu_xFe_yS_z$ system ($CuFeS_2$, $Cu_5FeS_4$, $CuFe_2S_3$ etc.) are intensively studied by numerous experimental and theoretical methods. This is due to a wide range of chalcogenide properties and, as a consequence, a wide potential area of practical application of chalcogenides, including as a source of copper metal in mining industry.

The valence states of transition metal Cu and Fe atoms [1, 2], the nature of phase transition at about 50 K and the magnetic structure of chalcopyrite $CuFeS_2$ below this temperature [3, 4] still remain open questions. Currently, there is no consensus on these issues despite extensive investigations over several decades.

Usually, experimental data for $Cu_xFe_yS_z$ system is analyzed on the basis of simple conception of oxidation states: two valence states for copper ($Cu^{1+}$ and $Cu^{2+}$), iron ($Fe^{3+}$ and $Fe^{2+}$) and mixed states. As a result, two probable ionic states $Cu^{1+}Fe^{3+}S_2^{2-}$ and $Cu^{2+}Fe^{2+}S_2^{2-}$ of chalcopyrite could exist. The monovalent copper has a completely occupied $d$-shell ($3d^{10}$ electron configuration, $^0S$ state), whereas divalent copper has one hole in $d$-shell ($3d^9$, $^2D$). The trivalent iron has half occupied $d$-shell ($3d^5$, $^6S$), and divalent iron has six $d$-electrons ($3d^6$, $^5D$).

Analysis of some X-ray absorption spectroscopy (XAS) and X-ray photoemission spectroscopy (XPS) data points to +1 oxidation state of copper in $CuFeS_2$. However, some of the data point to an admixture of the $3d^9$ configuration for the formally monovalent copper [5]. Cu

L-edge spectra have been interpreted in [6] as indicating a formal Cu valence between +1 and +2, where it was also noted that the $Cu^{2+}$ quantity is about a few percent.

The first neutron diffraction of chalcopyrite [7] has shown that the value of magnetic moment of Fe atoms (3.85 $\mu_B$) is reduced in comparison with 5 $\mu_B$ which is expected for $Fe^{3+}$ ion. At the same time, Cu magnetic moment is predicted to be about 0.2 $\mu_B$ or less. These results were confirmed later in [3] for synthetic chalcopyrite, and the magnetic moment of Cu was estimated to be approximately 0.05 $\mu_B$ (as opposed to nonmagnetic $Cu^{1+}$ state). However the latest high-resolution data [4] shows that introduction of a moment on the copper site gave no statistically meaningful improvement to the quality of fit to the data. At the same time, most of the theoretical and experimental investigations establish the covalent character of bonding in chalcopyrite.

The motivation of the present study is an electronic structure calculation aimed at highlighting several aspects of chalcopyrite $CuFeS_2$, including band structure, the oxidation state and magnetic moments of Cu and Fe atoms and their comparison with the available spectroscopic data.

**Crystal structure and details of calculations**

Chalcopyrite $CuFeS_2$ crystallizes in a body centered tetragonal Bravais lattice with space group $I\bar{4}2d$ ($D_{2d}^{12}$) and lattice parameters $a$=5.289 Å, $c$=10.423 Å [8]. Unit cell of chalcopyrite consists of four formula units (Fig. 1). Cu, Fe and S atoms occupy 4$a$ (0,0,0) and (0,1/2,1/4), 4$b$ (0,0,1/2) and (0,1/2,3/4), 8$d$ ($x$,1/4,1/8), (–$x$,3/4,1/8), (3/4, $x$,7/8) and (1/4,–$x$,7/8), $x$=0.2574 Wyckoff positions, respectively [9]. Unit cell of chalcopyrite has a double zincblende ZnS cell, in which the replacement of Zn ions by Cu and Fe elements takes place. Each atom of Cu and Fe is tetrahedrally coordinated by four S atoms and each atom of S is coordinated by two Cu and two Fe atoms. The $FeS_4$ tetrahedra have the perfect shape, whereas $CuS_4$ tetrahedrons are flattened in the $c$ direction of the cell ($c/a$=1.97) due to different S-Cu-S angles. Magnetic space group of chalcopyrite is also $I\bar{4}2d$ [7].

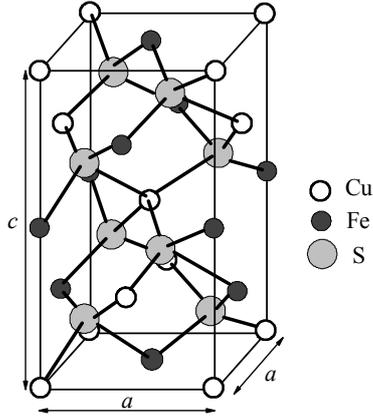

**Fig. 1**. Crystal structure of chalcopyrite $CuFeS_2$.

First-principles self-consistent-field spin-polarized electronic structure calculations were performed using Quantum Espresso package [10], which is based on the density functional theory (DFT). The local exchange-correlation potential was calculated by the local density approximation (LDA) using the scheme of Perdew–Zunger [11]. Electron-ions interactions were described by ultrasoft pseudopotentials generated according to a modified Rappe–Rabe–Kaxiras–Joannopoulos scheme [12]. The pseudopotentials were generated in the valence electron configurations $[3d^{10}4s^1]$ for the copper atom, $[3d^74s^1]$ for the iron and $[3s^23p^4]$ for the sulfur. Plane-wave basis set cut-offs for the wave functions and the augmented density were 60 and 400 Ry, respectively. The Kohn-Sham equations were solved numerically in a three-dimensional 6×6×6 Monkhorst-Pack grid [13]. 12×12×12 grid was used for calculations of electron density of states. The convergence threshold for self-consistent iterative procedure was $10^{-9}$ Ry.

**Results and discussion**

The calculated band structure and density of states are shown in Fig. 2 and Fig. 3. The Fermi level has been taken as the zero of the energy scale. Our first principle calculations in the frame of LDA approach predict that chalcopyrite has the conductive properties. While this has not been confirmed by experimental findings, it is in agreement with earlier studies [14]. At the same time, density of states at the Fermi level is small and 0.06 eV shift of the unfilled bands would lead to a zero-gap state. Thus we have good reasons to believe that LDA should predict appropriate valence state for chalcopyrite.

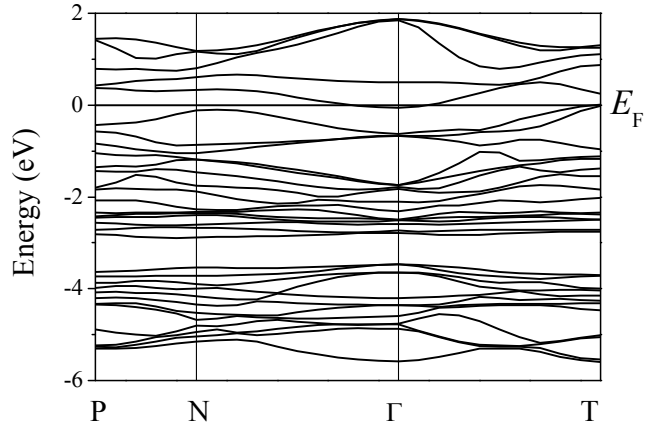

**Fig. 2**. The calculated band structure of chalcopyrite.

Our calculations predict strong mixing of Cu and Fe *d*-orbitals. The valence bands are composed mainly of the 3*s*, 3*p* orbitals of S and 3*d* orbitals of each Cu and Fe. The valence bands laid from -13.91 to -12.40 eV (not shown) come from the 3*s* orbitals of S. The bands in the energy range from -5.69 to -3.39 eV are composed mainly of 3*p* orbitals of S with admixing of 3*d* orbitals of Cu and Fe. The total number of states in this interval is 12, with 58.3 % from 3*p* orbitals of S, 15.6 % from 3*d* orbitals of Cu and 26.1 % from 3*d* orbitals of Fe. The last range from -2.97 eV to the Fermi level is composed mainly of 3*d* orbitals of Cu and Fe with admixture of 3*s*, 3*p* orbitals of S. The total number of states is 15 with 54.4 % from 3*d* orbitals of Cu, 24.0 % from 3*d* orbitals of Fe and 21.6 % from 3s, 3p orbitals of S.

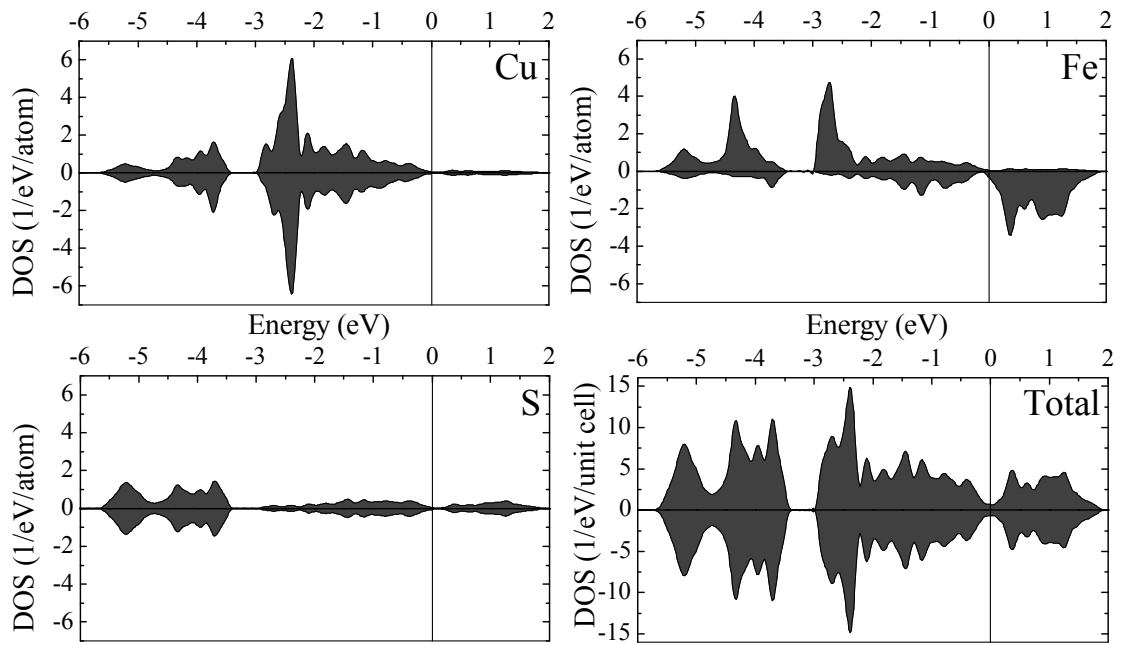

**Fig. 3**. Calculated partial and total density of states of chalcopyrite.

Results of population analysis are compared with some experimental and theoretical data in Table 1. It should be noted that the determination of absolute numbers of electrons for the electronic shells depends on the method of separation of electrons delocalized over the chemical bonds. In our case, Löwdin population analysis was applied [18]. As can be seen in Table 1, calculated numbers $N_{3d}$ of $3d$ electrons for transition metal atoms are in agreement with the published data. As noted in [1], the intensity of the satellite in the XPS spectra is too small to give a precise value of $N_{3d}$ for copper. Taking into account the value of $N_{3d}$=9.69 for copper, it seems reasonable to select the oxidation state as $Cu^{1+}Fe^{3+}S_2^{2-}$. Nevertheless, the calculated effective atomic charges for Cu and Fe are +0.73 and +0.76, respectively.

Table 1. The number of $3d$ electrons

| $N_{3d}$ | Experiment [5] | Theory | | | |
|---|---|---|---|---|---|
| | | [15] | [16] | [17] | This study |
| Cu | >9.5 | 9.7 | 9.63 | 9.75 | 9.69 |
| Fe | 4.6±0.1 | 6.8 | 6.12 | 6.86 | 6.68 |

Calculated magnetic moment of iron 3.29 $\mu_B$ is in agreement with the values found from neutron diffraction experiments (3.85 $\mu_B$ [7], 3.80 $\mu_B$ [5], 3.42 $\mu_B$ [3], 3.57 $\mu_B$ [4]). Very small ~$10^{-3}$ $\mu_B$ or vanishing magnetic moment was predicted for copper atom. Notably, recent neutron diffraction studies were unable to show unambiguously the presence of small Cu magnetic moments, although such possibility cannot be excluded at this stage [4].

As discussed above, LDA approach allows to describe the formal oxidation state for chalcopyrite rather well. This analysis is in agreement with some experimental XPS and XAS results [14]. At the same time, our and previous [14] LDA approach do not provide adequate description of conductive properties. It would be beneficial to compare the calculation results with other spectroscopic data provided, for example by Mössbauer effect or/and nuclear magnetic and quadrupole resonance (NMR/NQR). As was shown, for instance, for covellite (CuS) [19], the application of NQR/NMR spectroscopy can be useful to solve the problem of intermediate valence for copper sulfides. Moreover, application of generalized gradient approximation (GGA) together with Mössbauer/NMR data appears to be potentially promising due to the advantages of these methods for studies of magnetically ordered states. Preliminary comparative investigations of chalcopyrite using a combination of NMR and Mössbauer spectroscopy have been carried out at room temperature [20], however studies at lower temperatures remain to be done.

**Conclusion**

Our first principle calculations prove that LDA is partly efficient for explanation of properties of chalcopyrite $CuFeS_2$. However, LDA approach is ineffective for describing the conductive properties of chalcopyrite. At the same time, LDA is suitable for investigation of magnetic and oxidation states of $CuFeS_2$.

The Löwdin population analysis employed in this study allows us to conclude that the formal valence state of chalcopyrite is $Cu^{1+}Fe^{3+}S_2^{2-}$, however the real valence state of Cu is neither monovalent $Cu^{1+}$ nor divalent $Cu^{2+}$ due to the covalence effects. Implementation of other spectroscopic methods (Mössbauer, NMR and other) together with theoretical calculations especially at low temperatures would be expedient for elucidation of the problems noted.

**Acknowledgements**

This work was partly supported by RFBR grant under No. 12-02-31282 (mol_a).